# Collective excitations in liquid DMSO : FIR spectrum, Low frequency vibrational density of states and ultrafast dipolar solvation dynamics

Milan Hazra and Biman Bagchi[*]

SSCU, Indian Institute of Science, Bangalore 560012, India

## Abstract

Valuable dynamical and structural information about neat liquid DMSO at ambient conditions can be obtained through study of low frequency vibrations in the far infrared (FIR), that is, terahertz regime. For DMSO, collective excitations as well as single molecule stretches and bends have been measured by different kinds of experiments such as OHD-RIKES and terahertz spectroscopy. In the present work we investigate the intermolecular vibrational spectrum of DMSO through three different computational techniques namely (i) the far-infra red spectrum obtained through Fourier transform of total dipole moment auto time correlation function, (ii) from Fourier transform of the translational and angular velocity time autocorrelation functions and a (iii) quenched normal mode analysis of the parent liquid at 300K. The three spectrum, although exhibit differences among each other, reveal similar features which are in good, semi-quantitative, agreement with experimental results. Study of participation ratio of the density of states obtained from normal mode analysis shows that the broad spectrum around 100 cm$^{-1}$ involves collective oscillations of 300-400 molecules. Dipolar solvation dynamics exhibit ultrafast energy relaxation (dipolar solvation dynamics) with initial time correlation function around 140 fs which can be attributed to the coupling to the collective excitations. We compare properties of DMSO with those of water vis-a-vis the existence of the low frequency collective modes. Lastly, we find that the collective excitation spectrum exhibits strong temperature dependence.

## I. Introduction

In two seminal papers, Zwanzig [1,2] investigated the existence and utility of collective dynamical modes in the description of liquid state dynamics. The basic idea of his approach was as follows. If a collective mode is to provide a useful description of dynamics, it should have a long lifetime. If L is the Liouville operator, then the the condition that $A$ is a collective mode is given by

$$\frac{dA}{dt} = LA \simeq i\, zA \qquad (1)$$

Where $z$ is the eigenvalue of the Liouville operator L. Usually z is a complex number $z = \omega + i\varepsilon$, where $\omega$ gives the frequency of the collective mode and $\varepsilon$ is its life time. For a "good" collective excitation, the lifetime of the mode must be long, that is $\varepsilon$ should be small.

Such excitations are well-known in plasma fluids where it is known as plasmons. [3] In analogy, Lobo et al studied the existence of such excitations in dipolar liquids. They named such possible excitations as dipolarons.[4] They were also investigated by Pollack and Alder via computer simulations [5,6] Possible existence of such excitations was investigated by Chandra and Bagchi by using a molecular theory. [7] Although oscillations in collective longitudinal polarization was observed, the lifetimes were found to be too short lived to assign them as collective excitations. In dense liquids, one finds propagating shear waves that can be regarded as collective excitations of liquids. Actually, existence of such shear wave excitations motivated Zwanzig to look for collective excitations. It is interesting to note that Zwanzig's own conclusion as to the existence of long-lived collective excitation in liquids was also negative.

Since then a host of experimental studies that employed non-linear spectrscopic techniques have detected collective excitations in the low frequency domain, like in the 25-100 cm$^{-1}$ range. The most well-known is liquid water where a 50 cm$^{-1}$ mode has been attributed to hindred translation, and oxygen –oxygen-oxygen (O – O –O ) bending mode.[8] It should be pointed out that even though short lived, these low frequency vibrational excitations in lqiuid water play important role in solvation dynamics and electron transfer reactions.[9,10,11]

Such modes are usually seen in the far infra-red part of the spectrum which are somewhat difficult to detect. These are currently being detected in increasing number by terahertz spectroscopy, Kerr relaxation and non-linear spectroscopic techniques.[12-20]

Recently, Castner et al. reported A Kerr relaxation study of dimethy sulfoxide (DMSO) where they observed the existence of broad and prominent low frequency response.[21] The same was simulated by Skaf and coworkers who also observed a broad low frequency peak in the far infra red (FIR) region, in the range of 50-150 cm$^{-1}$.[22]

Note that the existence of low frequency intermolecular vibrational excitations are of importance in the dynamical properties of the system. They would also play important role in the glass transition behavior at low temperature.

In addition to FIR spectroscopy (that measures rotational time correlation function), one can have access to the low frequency vibrational density of states from study of quenched normal mode of ths system..

Yet another method to obtain the vibrational density of states is to use the Fourier transform of velocity time correlation function. This latter method gives information close to the DOS obtained from QNM.[23]

Two important issues that are important to understand are (i) the number of molecules involved in these low frequency vibrational excitations, (ii) the lifetime of these excitations. The first one can be answered theoretically by studying the participation ratio of these modes but the second question of lifetime is quite difficult to answer. Note that the width of a lineshape contains both the density of states and the lifetime of the vibrational modes. It is difficult to untangle these two factors.

Fortunately, the density of states from QNM gives the pure density of states unadulterated by the lifetime.

The role of these collective excitations can be understood from the following expression for the time correlation function of a dynamical quantity X(t)

$$<X(0)X(t)> = \int d\omega \, Cos(\omega t) \, e^{-t/\tau} \, g(\omega) \frac{<X(0)^2>}{-\omega^2 + \omega_0^2 + \Gamma(\omega)}. \qquad (2)$$

In the above equation, $\omega_0$ is the harmonic harmonic frquency of a given mode, g($\omega$) is the density of states, $\Gamma(\omega)$ frequency dependent friction or damping coefficient and $\tau$ is the lifetime of the mode with frequency $\omega$.

Such an expression for time correlation function has been used with success to model solvation time correlation function and also response function probed in non-linear optical response measurements.

When the decay takes place in the ultrafast time scale, then the dynamics of the time correlation function is controlled by the vibrational modes, and interference between them.

The organization of the rest of the paper is as follows. In the next section we present simulation calculation of the far infra-red absorption spectrum from the total dipole moment auto time correlation function. We also present the Cole-Cole plot. In section III we present the power spectrum obtained from the Fourier transform of the velocity time correlation function. In section IV we portray the rotational density of states from space fixed angular velocity auto correlation function. In section V and VI we present our calculation and analysis of the vibrational modes from quanched normal mode study. In section VII we present the calculation of the participation ratio of these modes. In section VIII we discuss the temperature dependence of the density of states. Section IX incorporates the ultrafast solvation dynamics. Section X and XI concludes with a brief comparison to experiments and discussion.

## II. Far Infrared spectrum of neat DMSO

Far infra red spectrum of DMSO is related to the fourier transform of total dipole moment correlation function through linear response theory.

$$I(\omega) = \frac{1}{2\pi} \int_{-\infty}^{\infty} dt e^{-i\omega t} <M(0)M(t)>  \quad (3)$$

Where total dipole moment $M(t)$ of the system is the sum of the individual dipole moments $\mu_i(t)$ of DMSO molecules at time t.

$$M(t) = \sum_{i=1}^{N} \mu_i(t) \quad (4)$$

$I(\omega)$ represents absorption lineshape and $\omega$ is in frequency units. $\delta M(t)$ is fluctuation of dipole moment at time t from the avarage dipole moment.

While using the above relation in usual molecular dynamics simulation, problems arise due to periodic boundary condition as particles are leaving the box from one side and re enters from another. This leads to discontinuity in variation of coordinates. One solution to this problem is that velocity is continous and that leads us to the final expression of absorbtion lineshape which is related to auto correlation of time derivative of the total dipole moment of the system.[24-29] The final expression is written as

$$I(\omega) = \frac{1}{2\pi\omega^2} \int_{-\infty}^{\infty} dt e^{-i\omega t} \left\langle \frac{d\vec{M}(0)}{dt} \frac{d\vec{M}(t)}{dt} \right\rangle \quad (5)$$

It is fairly easy to show the equivalence of **Eq.5** to that of **Eq.3**. All the time correlation functions have the following property

$$\left\langle \frac{dM(0)}{dt} \frac{dM(t)}{dt} \right\rangle = -\frac{d^2}{dt^2} \left\langle M(0)M(t) \right\rangle \quad (6)$$

As only the real part of Fourier transform of dipole moment correlation function is related to IR absorption intensity, using **Eq.(6)** in R.H.S of real part of **Eq.(5)** and integrating by parts twice we regain **Eq.(3)**

$$\int_{-\infty}^{\infty} dt \cos wt \left\langle \frac{d\vec{M}(0)}{dt} \frac{d\vec{M}(t)}{dt} \right\rangle$$

$$= -\int_{-\infty}^{\infty} \cos \omega t \frac{d^2}{dt^2} \langle M(0)M(t) \rangle$$

$$= \omega^2 \int_{-\infty}^{\infty} \cos \omega t \langle M(0)M(t) \rangle \propto \omega^2 I(\omega)$$

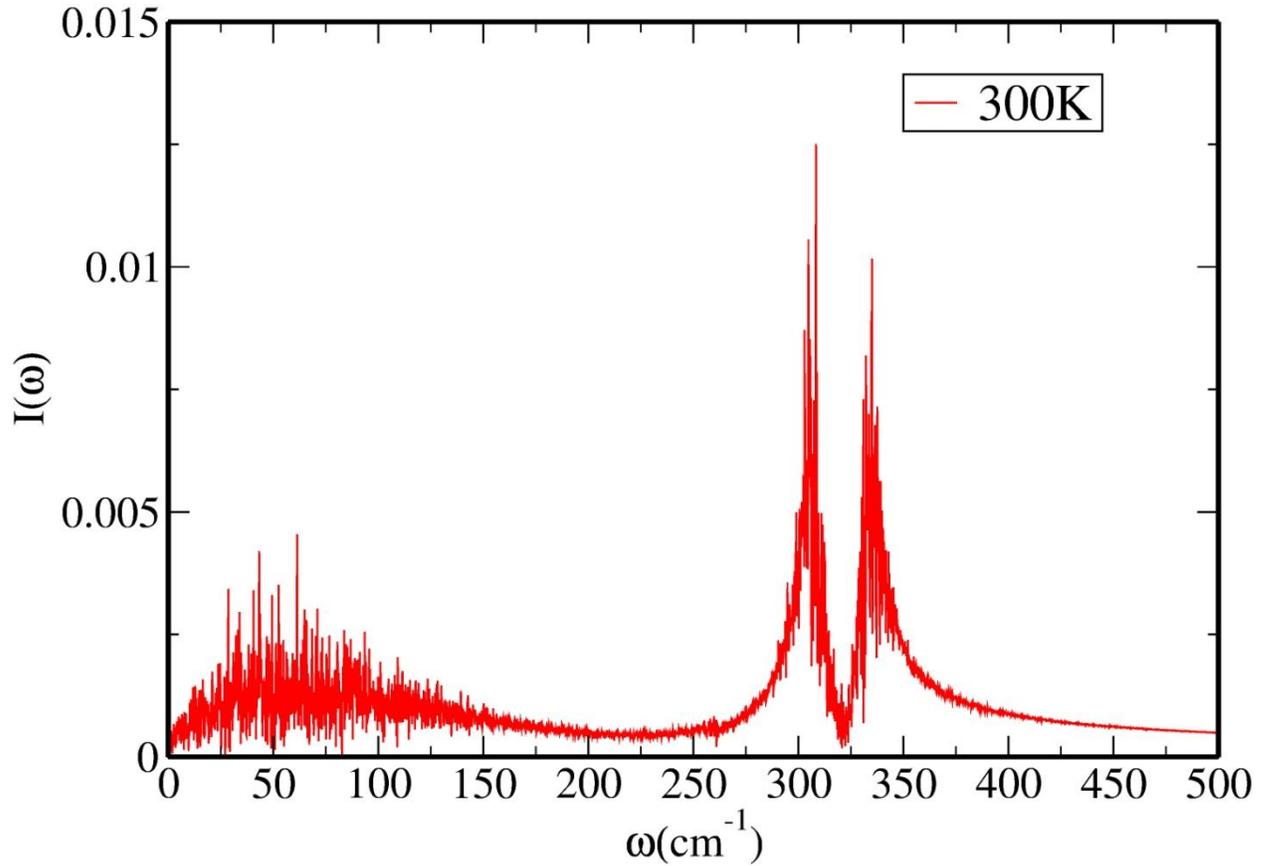

**Fig.1. The far infra red (FIR) spectrum for neat DMSO molecules at temparature 335K obtained from Eq.5 . Far infrared spectrum of DMSO incorporates both intermolecular and intra molecular osciliations. The broad peak in the frequency regime 0 to 150cm$^{-1}$ corrosponds to the many body osciliations present in neat DMSO Unlike water in DMSO both**

In **Figure 1** we present the far infra red (FIR) spectrum for neat DMSO molecules at temparature 335K obtained from **Eq.5**. Far infrared spectrum of DMSO incorporates both intermolecular and intra molecular osciliations. The broad peak in the frequency regime 0 to 150cm$^{-1}$ corrosponds to the many body osciliations present in neat DMSO. Unlike water in DMSO both intermolecular and intramolecular vibrations are present in the FIR frequency domain. With our flexible DMSO model we have captured the intramolecular vibrations as well. We find that the intramolecular modes have relatively sharp peak at 300cm$^{-1}$ and 350cm$^{-1}$. There is a small shoulder at 250cm$^{-1}$ confirmed through the single molecule far infra red spectra calculated from single molecular dipole orientational correlation and from power spectrum of both translational and rotational velocity components.

In dielectric relaxation, on the other hand, one is interested in frequency dependent dielectric constant. Frequency dependent dielectric constant is related to the collective dipole dipole autocorrelation of the system through the following relation

$$\varepsilon(\omega) - 1 = \frac{4\pi}{3k_B TV} \int_0^\infty e^{i\omega t} \left[ \frac{d}{dt} \langle M(0)M(t) \rangle \right] \qquad (7)$$

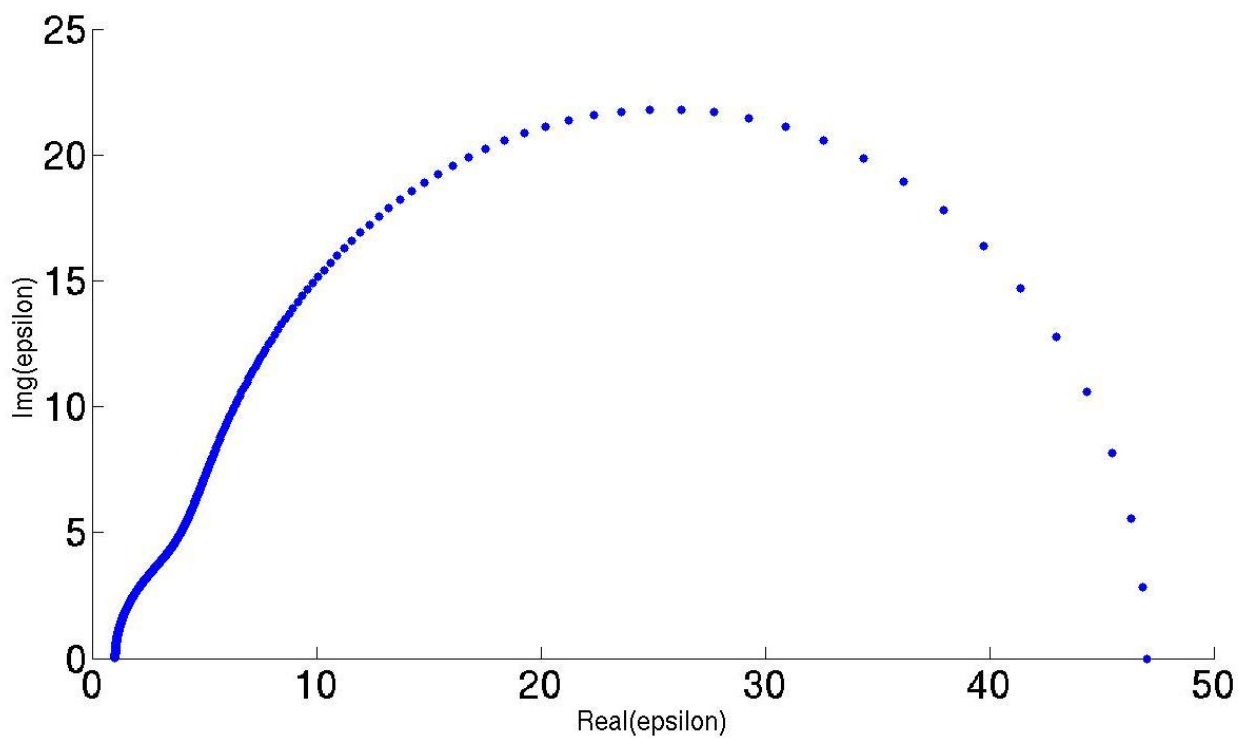

**Fig.2. The calculated Cole-Cole plot of frequency dependent dielectric constant of liquid DMSO at 300K. Imaginary part of dielectric constant is plotted against the real part. The collective dipole moment correlation function is biexponential in nature. When Real(epsilon) is in the range 1 to 6 ( in the frequency range that is, 1 to 4 THz), the collective osciliations are the reason to have a deviated semi circular cole cole plot.**

**Figure 2** represents Cole-Cole plot of frequency dependent dielectric constant at 300K for neat DMSO where the imaginary part of dielectric constant is plotted with respect to real part .Collective dipole moment correlation function is biexponential in nature. When Real(epsilon) is in the range 1 to 6 (in the frequency range 1 to 4 THz) the collective osciliations are the reason to have a deviated semi circular cole cole plot. The skewed appearance on the high frequency side suggests more than one relaxation process, specifically on the high frequency side of the spectrum. As discussed here, the ultrafast component arises from the collective oscillations shown in far infra red spectrum in the frequency range 0-100cm$^{-1}$.

## III. Power spectrum from translational velocity autocorrelation function

We represent in **Figure 3** the power spectrum obtained from translational velocity velocity autocorrelation function. Power spectra is related to VACF as

$$g(\omega) = \int_0^\infty dt\, e^{-iwt} \frac{<v(0)v(t)>}{<v(0)v(0)>} \qquad (8)$$

We observe a broad band in the frequency range 0 to 150cm$^{-1}$ and three sharp bands for methyl and sulfur atoms but two such sharp bands for oxygen in frequency range 225-400cm$^{-1}$. The broad band and the sharp bands are separated by a deep at 225cm$^{-1}$.

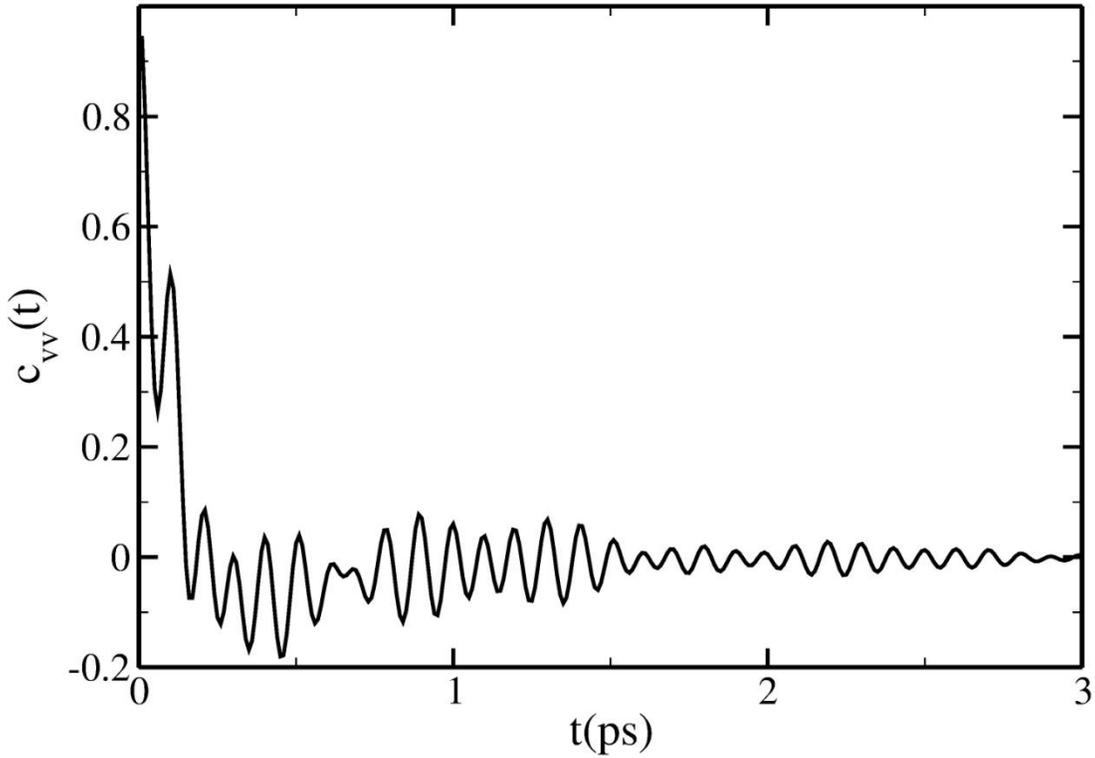

(a)

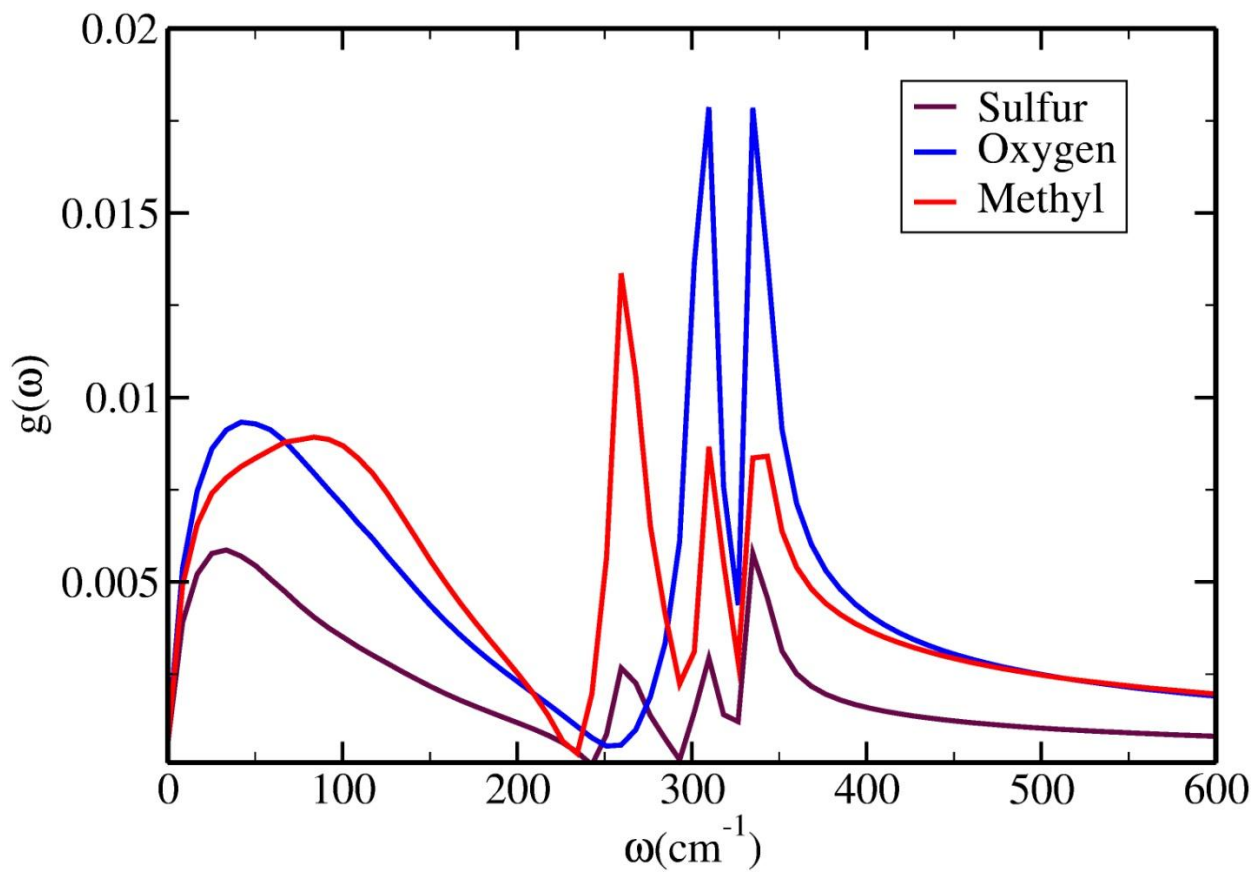

**(b)**

**Fig.3. (a) The calculated translational velocity autocorrelation of sulfur atoms of DMSO at 300K. (b) Vibrational power spectrum from translational velocity velocity autocorrelation in DMSO. A broad band in the frequency range 0 to 150 cm$^{-1}$ centred around 50 cm$^{-1}$ and three sharp bands for methyl and sulfur atoms but two such sharp bands for oxygen in frequency range 225-400 cm$^{-1}$. The broad band and the sharp bands are separated by a deep at 225 cm$^{-1}$. Unlike water DMSO has both inter and intra molecular vibrations in the FIR spectrum.**

**Figure 3(a)** shows the translational velocity velocity time autocorrelation of S atoms of neat DMSO at 335K. A broad band that is centred at 50 cm$^{-1}$ but spread between 0 to 150 cm$^{-1}$ and three sharp bands for methyl and sulfur atoms but two such sharp bands for oxygen in frequency range 225-400cm$^{-1}$. The broad band and the sharp bands are separated by a deep at 225cm$^{-1}$. *Unlike water DMSO has both inter and intra molecular vibrations in FIR spectrum.*

## IV. Power spectrum from angular velocity autocorrelation function

FIR is more coupled to rotation than translational motion. In order to isolate the rotational contribution to the density of states we calculate the angular velocity autocorrelation function of S=O bond.

Angular velocities in space fixed frame of S=O bonds are related to the euler angles ($\phi, \theta, \psi$) of the molecule in the following way [30]

$$\omega_x = \dot{\psi} \cos\theta \cos\phi - \dot{\theta} \sin\phi$$
$$\omega_y = \dot{\phi} - \dot{\psi} \sin\theta \tag{9}$$
$$\omega_z = \dot{\psi} \cos\theta \sin\phi - \dot{\theta} \cos\phi$$

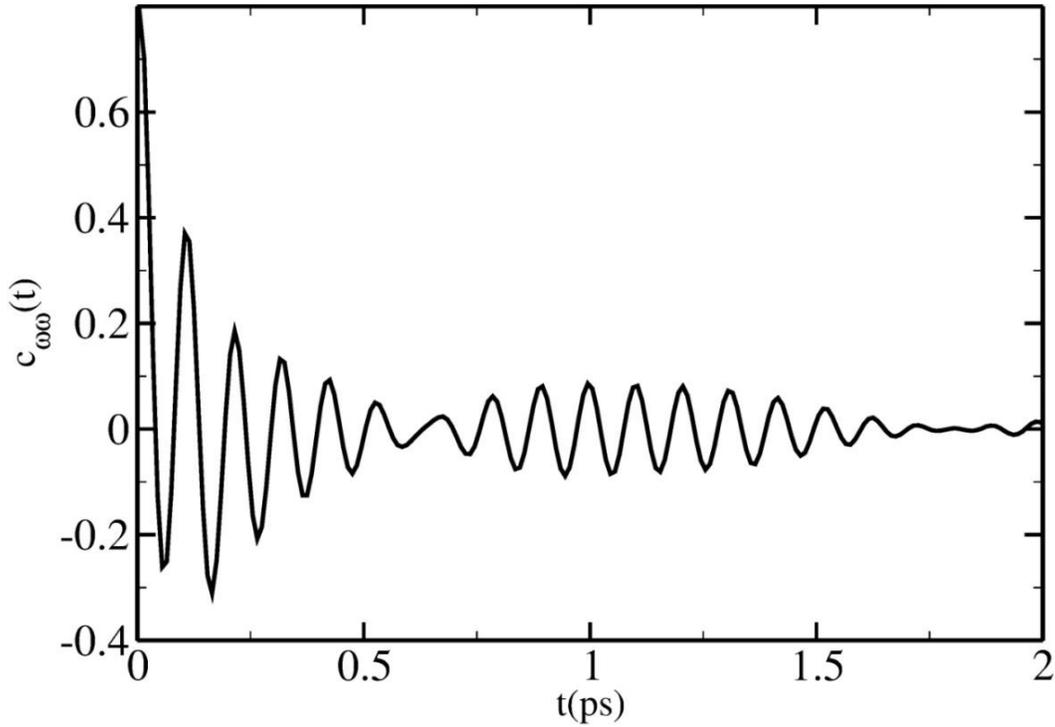

(a)

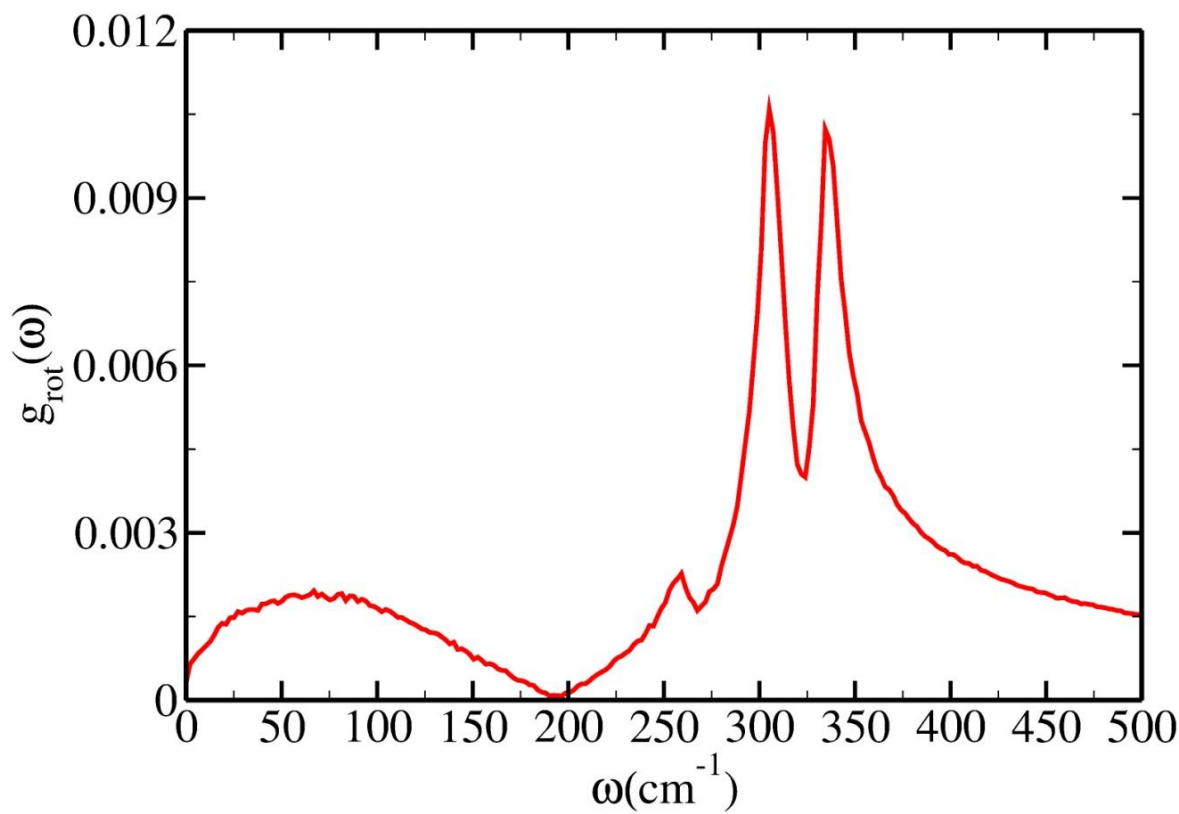

**Fig.4 (a) The angular velocity autocorrelation function of S=O bond in space fixed frame .(b) The corrosponding rotational density of states of the molecule. In contrast to the translational density of states, the rotational density of states are in fair agreement to the far infra red spectra.**

We observe highly osciliatory behavior in the rotational velocity auto correlation function in the time span of 500fs. This can be attributed to the collective osciliations of liquid DMSO. This time span can be regarded as the lifetime of the collective modes. The calculated rotational density of states of the molecule are in fair agreement to the far infra red spectra.

## V. Calculation of quenched normal modes

Quenched normal mode technique has been developed as a numerical tool to study the vibrational spectrum for disordered systems. In general, the density of states from the quenched normal mode study can be exactly related to the vibrational spectrum observed through experiments. On the other hand inverse participation ratio is useful to study localization of the normal modes obtained. The spectrum of flexible DMSO model is well separated through a deep in two frequency domains, low frequency regime (0-150cm$^{-1}$) corresponds to collective movement and high frequency (250-350cm$^-$1) represents intra-molecular modes. Instead of molecular point of view we look into the picture from atomic level in which atoms are regarded as the basic units.

The usual constant pressure and constant temparature molecular dynamics simulation is first used to equilibriate a system of 960 DMSO molecules at a temparature of 300K for 2 ns. Nose-Hoover thermostat is used to keep the temparature fixed. A constant pressure is maintained through Parinello-Rahman barostat. The simulation is carried out with Gromos 53a6 forcefield and with no constraints. We do not include methyl hydrogens in our simulation. The intramolecular bonds has been modeled as simple harmonic osciliators with a specific force constant. In **Table.1** we show the force constants involved in stretching, bending vibrations.

$$E_{bond}(r) = \frac{1}{2} k_{bond} (r - r_0)^2. \qquad (10)$$

$$E_{angle}(r) = \frac{1}{2} k_\theta (\theta - \theta_0)^2. \qquad (11)$$

Where $r_0$ is equilibrium bond length and $\theta_0$ implies equilibrium angles. $k_{bond}$ and $k_\theta$ represent the force constants of various stretches and bends respectively in the molecule.

**Table.1** shows the equilibrium bond lengths, angles and force constants for various bonds and angles involved during stretches and bends respectively.

**Table 1: Different stretch and bend parameters for our model DMSO.**

| stretch or bend type | Force constant ($k_{bond}$ or $k_\theta$) (KJ/mol/nm$^2$ for stretching, (for bending motion KJ/mol/rad$^2$) | Equilibrium Bond length and angle |
|---|---|---|
| S-C | 376560 | 0.195nm |
| S=O | 502080 | 0.153nm |
| O-S-C | 460.240 | 106.75$^0$ |
| C-S-C | 460.240 | 97.40$^0$ |

The equilibriated liquid configuration is energy minimized first with conjugate gradient then with L-BFGS algorithm to obtain the minimum of potential energy. The final energy minimized structure is used calculate $12N \times 12N$ dynamical matrix of second derivative of potential energy with respect to mass weighted atomic coordinates of the atoms in the system. Diagonalization of the hessian yields 12N eigenvecors and their respective frequencies. For a particular configuration of a liquid we calculate the density of states by binning the eigenfrequencies which are obtained from diagonalizing the matrix. Density of states has been avaraged over 5 independent configurations within 50 ps time window.

## VI. Density of states from quenched Normal modes

**Figure 5** shows the computed density of states from the quenched normal mode analysis. Low frequency modes from 0 to 150ps involve intermolecular coupled vibrations. High frequency sharp peaks involve s intra molecular modes. There is a deep from 150cm$^{-1}$ to 250cm$^{-1}$ through which the inter and intra molecular vibrations are well separated.

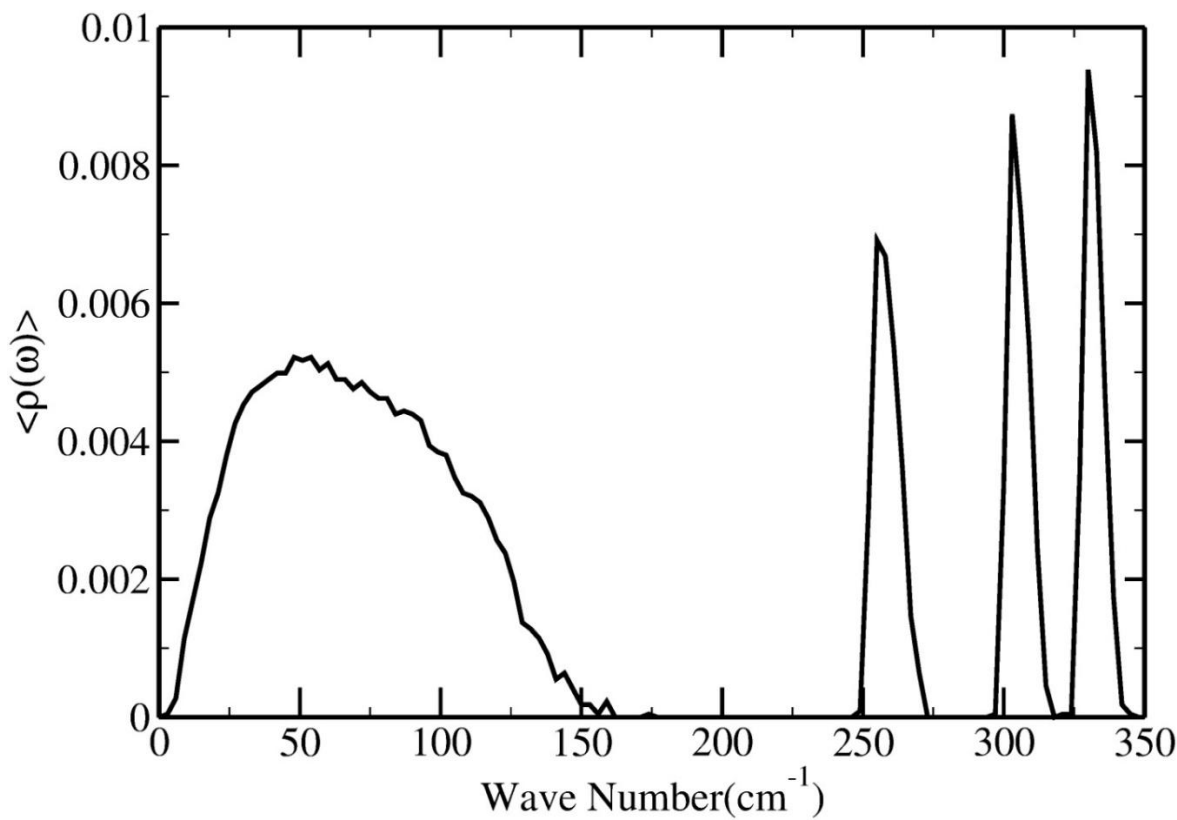

**Fig.5.Density of states at low frequency calculated from quenched normal modes of DMSO at temperature 300K. Far infra red spectrum of DMSO consists both intermolecular and intra molecular modes. Broad peak from 0-150 cm$^{-1}$ but centred around 50 cm$^{-1}$, shows the collective intermolecular oscillations that involves the in plane twist motion of the molecules while the sharp peaks at 255cm$^{-1}$ , 310cm$^{-1}$ , 330cm$^{-1}$ involves intra molecular vibrational motion. These peaks involve motion of -CH$_3$ groups. The peak at 255cm$^{-1}$ correspond to intra-molecular methyl-methyl opening and closing. Peak at 310 cm$^{-1}$ and 330cm$^{-1}$ correspond to methyl-methyl out of plane twist and wagging.**

There is a dip from 150cm$^{-1}$ to 250cm$^{-1}$ through which the inter and intra molecular vibrations are well separated . The broad peak from 0-150 cm$^{-1}$ represents the collective intermolecular oscillations that involve the in plane twist motion of the molecules while the sharp peaks at 255 cm$^{-1}$ , 310cm$^{-1}$ , 330cm$^{-1}$ involves intra molecular motion. These peaks involve motion of -CH$_3$ groups. The peak at 255cm$^{-1}$ correspond to intra-molecular methyl-methyl opening and closing. Peak at 310 cm$^{-1}$ and 330cm$^{-1}$ correspond to methyl-methyl out of plane twist and wagging.

## VII. Localization and delocalization of modes

Since these quenched normal modes are excitations that we calculate from a frozen random liquid configuration, these modes are expected to be localized in space. On the other hand, these are low frequency modes that we expect to be at least partly delocalized. To understand the localization properties, we calculate the inverse participation ratio of these modes defined defined as

$$R^{II}{}_\alpha = \sum_j \left[ \sum_\mu (U_{\alpha,j\mu})^2 \right]^2 \qquad (12)$$

where $\alpha$ denotes the index of eigenvector and j signifies the index of the molecule $\mu$ denotes the intramolecular coordinate indices. For a completely localized normal mode in only one degree of freedom out of 3N degrees of freedom, only one eigenvector element has to be non zero. The inverse participation ratio (IPR) as a consequence has to be 1. The above defination in **Eq.12** thus gives the total fraction of molecules involved in a particular normal mode.

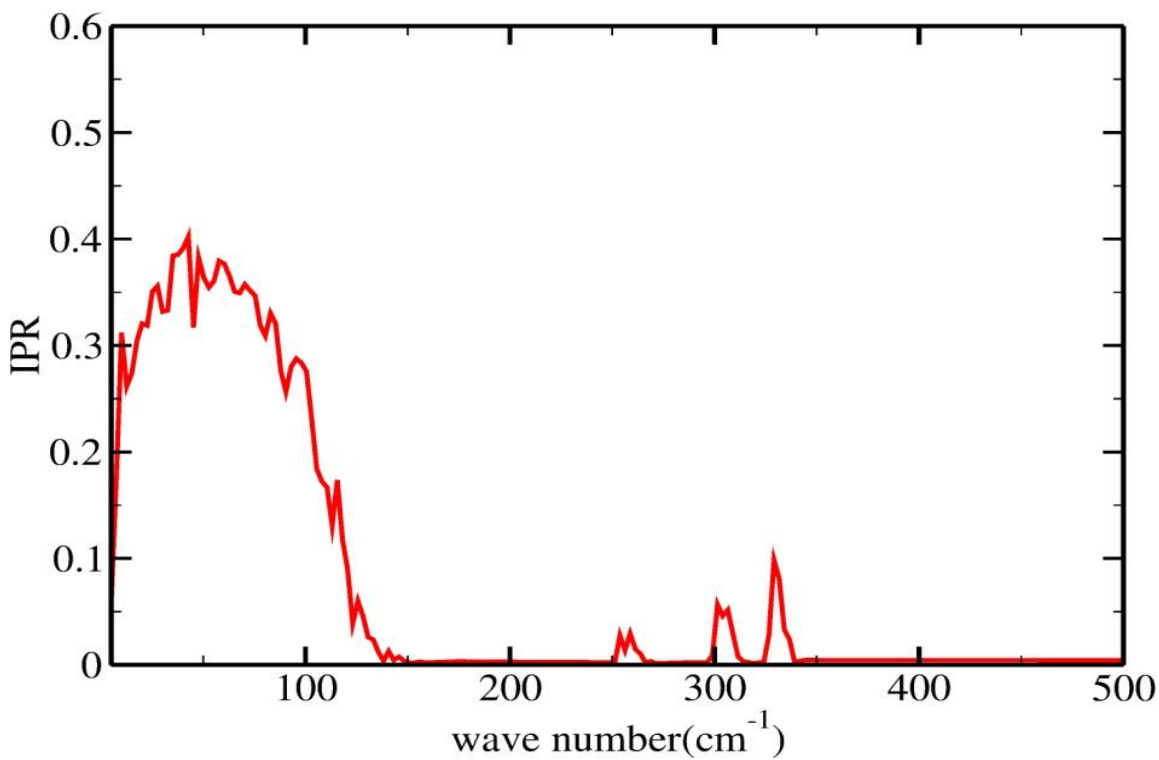

**Fig.6. Inverse participation ratio of DMSO molecules in each normal mode characterized by frequency in cm$^{-1}$. The normal modes around 50 cm$^{-1}$ are somewhat delocalized, involving almost 40% of the molecules of the system. Normal modes in frequency range 250cm$^{-1}$ to 350cm$^{-1}$ are localized – only a very small number of molecules are involved in vibration. However, somewhat surprisingly these intramolecular vibrations are not single molecular phenomena. These intramolecular motions involve around 20 , 50 , 100 molecules each for the 250cm$^{-1}$ ,310cm$^{-1}$ and 330cm$^{-1}$ respectively.**

**Figure 6** shows the inverse participation ratio of neat DMSO molecules at 300K. Therefore, this figure provides the answer to the question as to how many molecules are involved in each mode. For a delocalized collective vibrations inverse participation ratio is almost three to four times those of the intra molecular vibrations.

## VIII. Temperature dependence of FIR and power spectra

It is important to know how the far infra red spectrum as well as power spectrum of DMSO depends on temperature. We performed calculation at three different temparatures. In **Figure 7** surprisingly intermolecular modes show prominent temperature dependence with broadening of the peaks while intramolecular modes does not show such behavior.

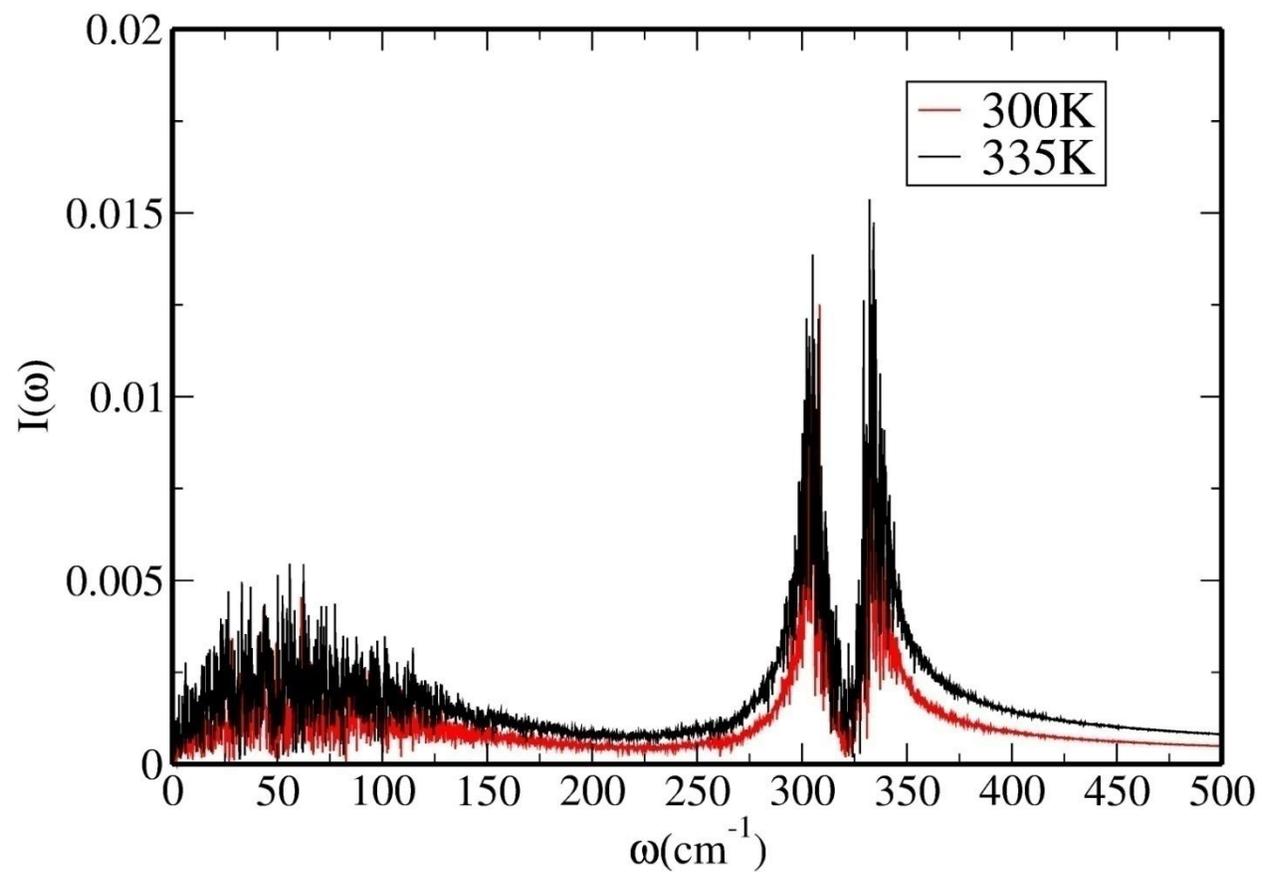

(a)

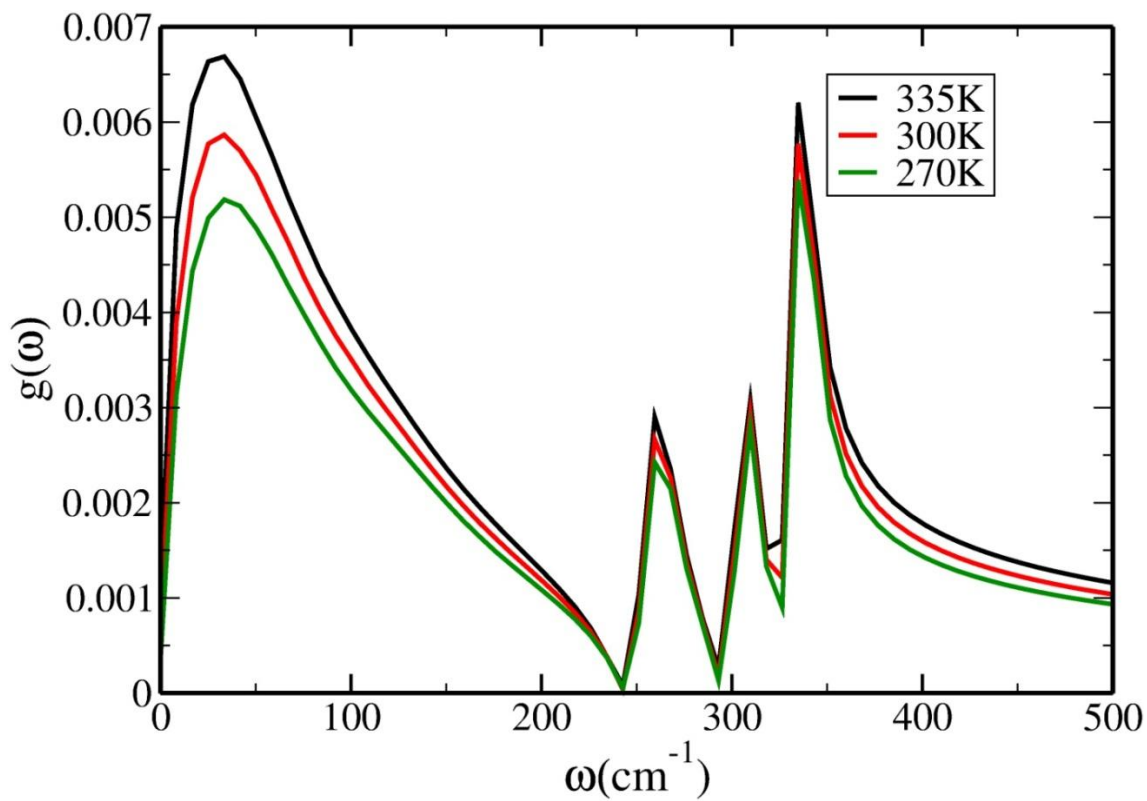

**(b)**

**Fig.7.(a) Temparature dependence of far infra red spectra at two different temperatures. (b) Temperature dependence of power spectrum of DMSO molecules calculated from velocity -velocity auto correlation function. Figure shows power spectrum at three different temperatures 335K , 300K and 270K. Collective vibrations show a prominent temperature dependence with increase and broadening of the peak intensity. No such effect is observed for intra molecular modes in power spectrum.**

**Figure 7** shows that temperature can have a strong influence on the line spectrum at low frequency. This is understandable because the collective mode depends on the thermodynamic state of the system.

## IX.  Solvation time correlation function

The solvation time correlation function $S(t)$ is usually defined as

$$S(t) = \frac{E_{solv}(t) - E_{solv}(\infty)}{E_{solv}(0) - E_{solv}(\infty)} \tag{13}$$

Where $E_{sol}(t)$ is the solvation energy of solute at time t. When the response is linear to external perturbation, the solvation time correlation function can be represented as the equilibrium energy-energy time correlation as the following [9,10,31]

$$S(t) = \frac{<\delta E_{sol}(0)\delta E_{sol}(t)>}{\langle \delta E_{sol}(0)^2 \rangle} \tag{14}$$

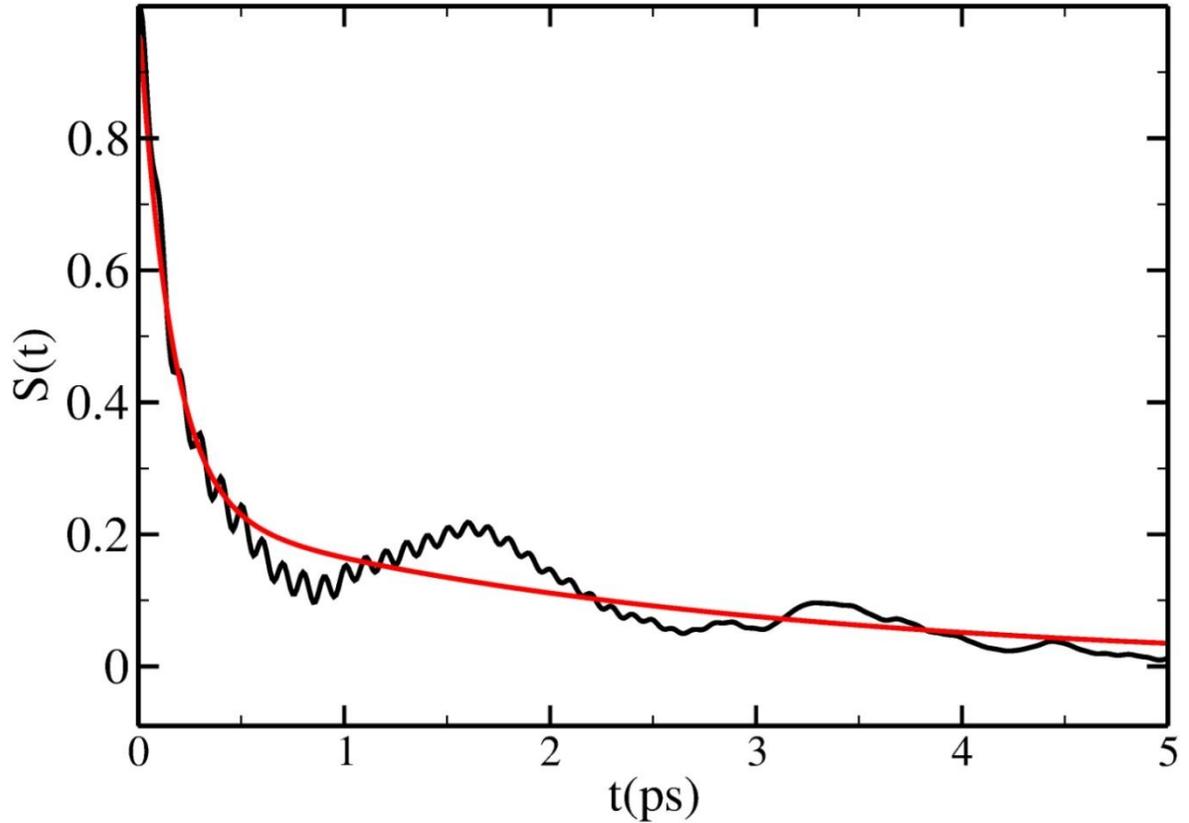

**Fig.8.** Solvation time correlation function in neat liquid DMSO. We find a bi exponential function as the best fit with an ultrafast component 157fs and a relatively slower component with 1.75ps. This ultrafast component gives the lifetime of the collective vibrations roughly which is of the order of 200fs.

Where $\delta E_{sol}(t)$ is the fluctuation in energy from equilibrium time average

$$\delta E_{sol}(t) = E_{sol}(t) - <E_{sol}> \quad (15)$$

**Figure 8** shows the salvation time correlation function in neat DMSO. Solvation time correlation function in neat liquid DMSO. We find a bi-exponential function of the form

$$S(t) = a_0 e^{-\frac{t}{\tau_1}} + (1-a_0)e^{-\frac{t}{\tau_2}} \quad (16)$$

as the best fit with an ultrafast component ($\tau_1$) 157 fs with amplitude $a_0 = 0.34$ and a relatively slower component ($\tau_2$) 1.75ps with amplitude 0.66. This ultrafast component gives the lifetime of the collective vibrations could roughly be of the order of 200 fs.

Both **Figure 8** and the calculated values of the relaxation times bear similarities to dipolar solvation dynamics of a polar probe in liquid water. In the case of water, we find an ultrafast Gaussian component with time constant between 50-100 fs with amplitude estimates ranging from 50 to 70% and a slower component with time constant of about 1 ps with amplitude about 20-30 %. In DMSO, the time constants and amplitudes differ but general pattern seems to remain the same.

The ultrafast component in water is attributed to the collective excitations between 50 to 600 $cm^{-1}$. In this case, we can attribute the ultrafast component to the collective intermolecular modes centred around 50 $cm^{-1}$.

## X. Comparison with experiments

The collective and single molecule vibrational density of states was measured by Castner et al. using optical heterodyne-detected Raman-induced Kerr effect spectroscopy for the entire range of composition at 294K. In the spectral resolution 0-750cm$^{-1}$, OHD-RIKES spectrum involves both collective and intra molecular modes. According to them low frequency intermolecular collective dynamics gives rise to a broad peak around 100cm$^{-1}$ which has been attributed to the librational motion of DMSO. They also concluded that modes at 303cm$^{-1}$, 333cm$^{-1}$, 383cm$^{-1}$ are due to intra molecular motion of DMSO.

The present studies seem to agree quite well with the experimental studies.

## XI. Conclusion

In this work we have employed different theoretical techniques to analyze the low frequency intermolecular vibrational modes. We find prominent existence of such modes around 50 $cm^{-1}$. When we calculate the power spectrum from linear velocity and angular velocity time correlation functions, we find the existence of these modes to figure prominently. The same modes are found in the FIR spectrum.

We observed that our results are in good agreement with known experimental studies of vibrational spectrum of liquid DMSO. Additionally, we found that while the intramolecular vibrational modes are localized on a few molecules, our modes centred about 50 $cm^{-1}$ are delocalized, with an involvement of about 400 DMSO molecules.

We have also observed that many of the dynamical features of liquid DMSO bear similarities to those found in liquid water.

In future studies, we shall focus on the elucidation of such dynamics in water-DMSO binary mixtures for which both theoretical and experimental results exist while no study of the vibration spectrum has yet been made.

To the best of our knowledge this study constitutes the first attempt correlating the density of states obtained from FIR and quenched normal mode with salvation dynamics and related phenomena in neat DMSO. Similar studies have been carried out by Cho et.al earlier.[23] We plan to explore the analogy further.

# ACKOWLDGEMENT

It is a pleasure to thank Professor K. Tominaga and his students of Kobe University, Japan for many illuminating discussions. We thank Sir J.C. Bose Fellowship for partial support of the work. We also acknowledge partial support from DST (India).